\documentclass[
10pt,
tightenlines,
showkeys,
nobibnotes,
nofootinbib,
aps,
superscriptaddress,
a4paper,
twocolumn,
]{revtex4-1}
\pdfoutput=1
\usepackage[T1]{fontenc}
\usepackage[latin9]{inputenc}
\usepackage{graphicx}
\usepackage{setspace}
\usepackage{amsmath}
\usepackage{amsthm}
\usepackage{enumerate}
\usepackage{textcomp}
\usepackage{amssymb}
\usepackage{relsize}
\usepackage{appendix}
\usepackage{babel}
\usepackage{amsfonts}
\usepackage{tabulary}
\usepackage{appendix}

\usepackage{rotating}
\usepackage{enumitem}
\usepackage{titlesec}
\usepackage{bm}


\begin{document}
\title{Reconfigurable Computation in Spiking Neural Networks}

\author{Fabio Schittler Neves}
\affiliation{Chair for Network Dynamics, Center for Advancing Electronics Dresden (cfaed), Dresden, Germany}
\affiliation{Institute for Theoretical Physics, TU Dresden, 01062 Dresden, Germany}
\email{fabio.neves@tu-dresden.de}

\author{Marc Timme}
\affiliation{Chair for Network Dynamics, Center for Advancing Electronics Dresden (cfaed), Dresden, Germany}
\affiliation{Institute for Theoretical Physics, TU Dresden, 01062 Dresden, Germany}
\affiliation{Cluster of Excellence Physics of Life, TU Dresden, 01062 Dresden, Germany}
\affiliation{Lakeside Labs, 9020 Klagenfurt am W{\"o}rthersee, Austria}
\email{marc.timme@tu-dresden.de}

\begin{abstract}
The computation of rank ordering plays a fundamental role in cognitive tasks and offers a basic building block for computing arbitrary digital functions. Spiking neural networks have been demonstrated to be capable of identifying the largest $k$ out of $N$ analog input signals through their collective nonlinear dynamics. By finding partial rank orderings, they perform $k$-winners-take-all computations. Yet, for any given study so far, the value of $k$ is fixed, often to $k$ equal one. Here we present a concept for spiking neural networks that are capable of (re)configurable computation by choosing $k$ via one global system parameter. The spiking network acts via pulse-suppression induced by inhibitory pulse-couplings. Couplings are proportional to each units' state variable (neuron voltage), constituting an uncommon but straightforward type of leaky integrate-and-fire neural network. The result of a computation is encoded as a stable periodic orbit with $k$ units spiking at some frequency and others at lower frequency or not at all. Orbit stability makes the resulting analog-to-digital computation robust to sufficiently small variations of both, parameters and signals. Moreover, the computation is completed quickly within a few spike emissions per neuron. These results indicate how reconfigurable $k$-winners-take-all computations may be implemented and effectively exploited in simple hardware relying only on basic dynamical units and spike interactions resembling simple current leakages to a common ground.
\end{abstract}


\maketitle

\section{Introduction} % SECTION 1: INTRO
\label{sec:introduction}
Rank ordering of signals plays a fundamental role in natural and artificial cognitive computations, in particular in attention-related tasks \cite{Bridewell2016theory, McKinstry2016Imagery,Rabinovich2013neural} where a small set of relevant information must be computed in real time from an array of sensory inputs. Natural environments present a continuous stream of concurrent analog signals that often are simultaneously time-dependent, multi-modal and high-dimensional. Yet such complex signals typically provide a basis for discrete decisions.
\begin{figure}[] 
\begin{centering}
\includegraphics[width=7.8cm,angle=0]{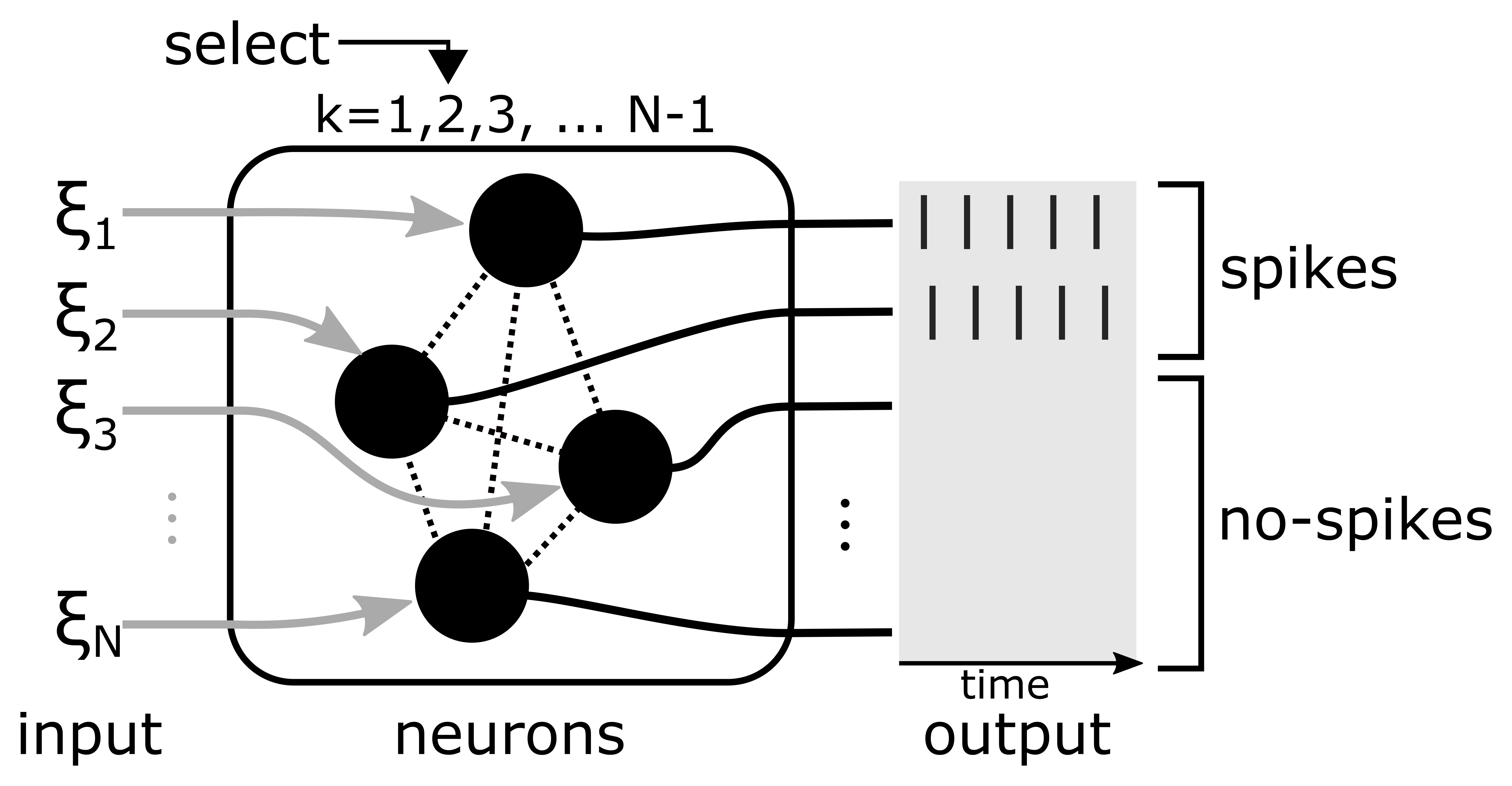}
\par\end{centering}
\caption[1st-entry]{ \small \textbf{Reconfigurable $k$-winners-take-all computation in a schematic network of spiking neurons.} Tuning one global parameter, the coupling strength $\epsilon$, selects $k$. A vector $\xi$, with $\xi_{1} > \xi_{2} > \xi_{3} > \dots > \xi_{N}$ (gray arrows) serves as an input to a network of spiking neurons (black circles) that dynamically computes a 2-WTA function of  the input vector $\xi$. Only those neurons receiving the largest two inputs send spikes whereas spikes from all other neurons are dynamically suppressed.}
\label{fig:intro}
\end{figure}
% Figure 1 ends here

Partial rank ordering, determining a subset of strongest signals by comparing a variety of inputs, offers a fundamental option for providing a discrete judgment. Finding the largest $k$ out of a set of $N$ analog inputs defines a $k$-winners-take-all ($k$-WTA) computation, which networks of spiking (formal) neurons have been shown to perform \cite{maass2000computation, Neves2012computation} via different approaches. While bio-inspired networks of pulse-coupled neurons can compute one winner-takes-all (1-WTA) functions via a combination of local excitation and longer range inhibition \cite{chen2017mechanisms}, e.g. bump states \cite{Laing2001stationary} or dynamical neural fields \cite{Sandamirskaya2014dynamics}, symmetrical networks of pulse-coupled neurons or phase-coupled oscillators may compute more general $k$-WTA functions exploiting complex periodic orbits akin to heteroclinic dynamics \cite{Krupa1997robust, rabinovich2001dynamical, Timme2002prevalence, Timme2003unstable, Ashwin2004encoding, Ashwin2005, Ashwin2005when, Neves2012computation, Wordsworth2008spatiotemporal}. Moreover, a recent study \cite{Wang2019wta} also provides an actual simple hardware implementation of a WTA neural-circuit with promising scalability via careful network design, also a mix of excitation and inhibition.

In all these studies, however, either  the number of  winners $k$ is fixed, hard to design, thus not easily (re)configurable, or computations typically take many spikes per unit or both \cite{rabinovich2001dynamical,  Timme2002prevalence, Timme2003unstable, Ashwin2005, rabinovich2006dynamical, Wordsworth2008spatiotemporal, Neves2012computation, Neves2017Noise}. Here, we propose a class of spiking neural network models that not only computes fast, in as few as $k$ spikes under ideal conditions (no noise and all starting from the same voltage), but  moreover provides intrinsic (re)configurability via changes to a single global parameter to perform $k$-WTA computations for different $k$, see Fig. \ref{fig:intro} for a schematic illustration.

We thus present a system that is directly (re)configurable by setting a single parameter to different values in order to directly switch between different well-defined computational tasks. Such reconfiguration is in stark contrast to learning processes in neural networks during which many parameters (coupling weights) evolve in a self-organized way to adapt the system to become capable of solving a given task.

\section{Spiking neural networks with proportional inhibition} % SECTION 2: SYSTEM
\label{sec:system}
We consider a network of $N$ spiking neurons that are oscillatory, i.e. send spikes periodically in the absence of coupling and external signals. The neurons' continuous-time dynamics is a singular type of integrate-and-fire neurons \cite{maass1999Pulse} and is characterized by a one-dimensional phase-like variable as in the generic model introduced in \cite{Mirollo1990synchronization}. Each neural unit $ i\in \{1, \ldots, N\}$ exhibits a voltage-like state variable $x_i$ satisfying the differential equations
\begin{equation}
	\frac{dx_{i}}{dt} = I_{i} -\gamma x_{i} + \xi_{i}(t) + \eta_i(t) + G(x_i, t),
	\label{eqn:voltageODE}
\end{equation}
where
\begin{equation}
	G(x_i, t) = \sum_{j=1}^{N} \sum_{t_{j,\ell} \in P_{j}} g(x_{i}) \delta(t-t_{j,\ell}),
	\label{eqn:voltageODE_2}
\end{equation}
for  $x_{i} \in [0, x^{\theta})$ where $x^{\theta}$ is a spiking threshold. The parameter $I_i$ represents a constant current, $\gamma$ is a dissipation parameter, $\xi_{i}$ an external signal serving as input and $\eta_i(t)$ represents an independent Gaussian white noise signal which strength is completely determined by its signal variance $\sigma^2$. Numerically, by definition, the noise contribution within any given small time-interval $\Delta t$ is a random real number with zero mean and variance $(\Delta t) \sigma^2$. Thus, one random number is generated at each integration interval and summed to the resulting voltage. To avoid numerically driven synchronization, noise is added at randomly generated times \cite{hansel1998numerical, Timme2003unstable, Klinglmayr2012Guaranteeing} (201 points per time unit per neuron) drawn independently from a Poisson distribution. Whenever a threshold is reached $x_{i} (t^-)=  x^{\theta}$, the state variable is reset to  $x_{i}(t) = 0$ and a pulse (spike) is sent to all other neurons, mathematically reflected in the time $t=t_{j,\ell}$ of the $\ell$th threshold crossing, $\ell \in \mathbb{Z}$, by the neuron $j$. Without loss of generality we fixed $x^{\theta}=1$. The sum in \eqref{eqn:voltageODE_2} is the contribution of all spikes arriving from the other $N-1$ neurons $j$ to $i$ at time $t$, where $P_{j}$ is the set of all times $t_{j,\ell}$ of spikes sent by neuron $j$. We propose to design the coupling function $g(x_j)$, also known as sensitivity function in 
the research field of coupled phase oscillators \cite{strogatz2000kuramoto,Winfree1967biological}, to be state dependent such that it is inhibitory with amplitude proportional to the voltage $x_i$ of the receiving neuron at times $t_{j,\ell}^-$, i.e.
\begin{equation}
\label{eq:connection}
	g(x_{i}) = - \epsilon x_{i}
\end{equation}
where $\epsilon<1$ is the coupling strength. As a consequence, the larger the voltage $x_i$, the larger the inhibitory effect of a received spike signal. To the best of our knowledge, this model setting has not been analysed or exploited for solving computational tasks (via spiking neural networks).
\begin{figure}[]
\begin{centering}
\hspace*{-0.3cm}\includegraphics[width=8cm,angle=0]{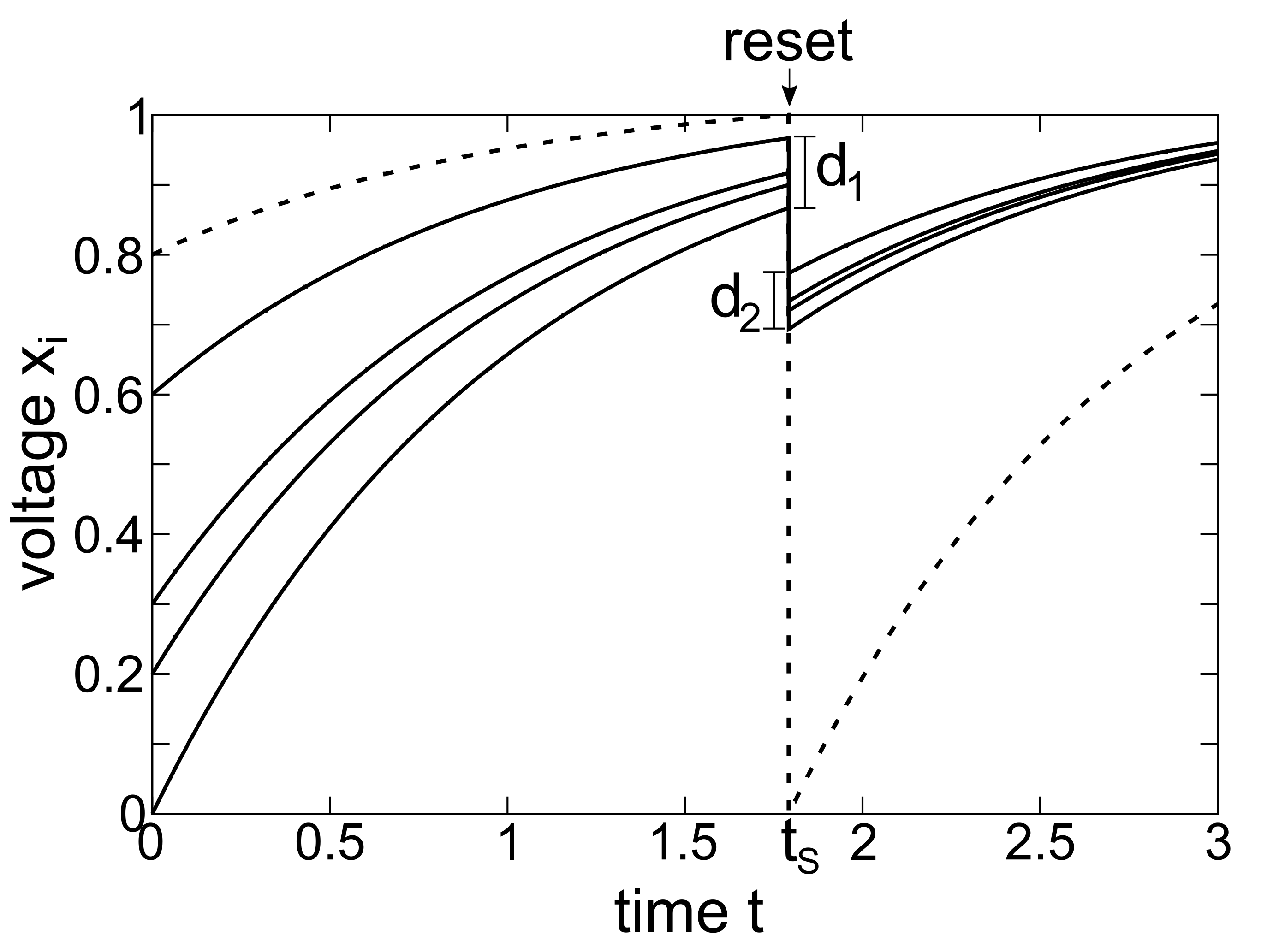}
\par\end{centering}
\caption[1st-entry]{ \small \textbf{Effect of inhibitory spike reception.} The continuous time evolution of state variables (voltages) $x_i(t)$ as a function of time $t$. Each line represents the voltage of one neuron $i$. When some neuron $j$ (dashed line) reaches the threshold, $x_j(t^-_s)=1$, it is reset, $x_j(t_s)=0$, and a spike is immediately sent to all other neurons. Because the spike response strength is proportional to the voltage of receiving neurons $x_i(t_s)=x_i(t^-_s) - \epsilon x_i(t^-_s)$, their voltages will be both decreased, $x_i(t_s)<x_i(t^-_s)$ and more strongly synchronized than before the spike reception, i.e.  $d_2 =  (1-\epsilon) d_1$ where $d(t)=\max_{i,i'\neq j}|x_i(t)-x_{i'}(t)|$, $d_1 = d(t_s^-)$ and $d_2 = d(t_s)$. Parameters for this illustration are $N=5$ and $\epsilon = 0.2$.}
\label{fig:compression}
\end{figure}
% Figure 2 ends here

For illustration, we here focus on networks with fixed parameters $\gamma = 1$ and  $I_i= 1.04=: I $. We vary the magnitudes of the input signals $\xi_i(t)$ and the coupling strength $\epsilon$ to select different $k$ of the $k$-WTA computation by achieving different collective dynamics. Furthermore, as concrete examples, we present networks of sizes $N=5$ and $N=50$ to emphasize different collective dynamics' aspects and the pulse-suppression mechanism's independence of the network size.

\section{$k$-winners-take-all via Pulse-Suppression} % SECTION 3: DYNAMICS
\begin{figure}[]
\begin{centering}
\includegraphics[width=8.2cm ,angle=0]{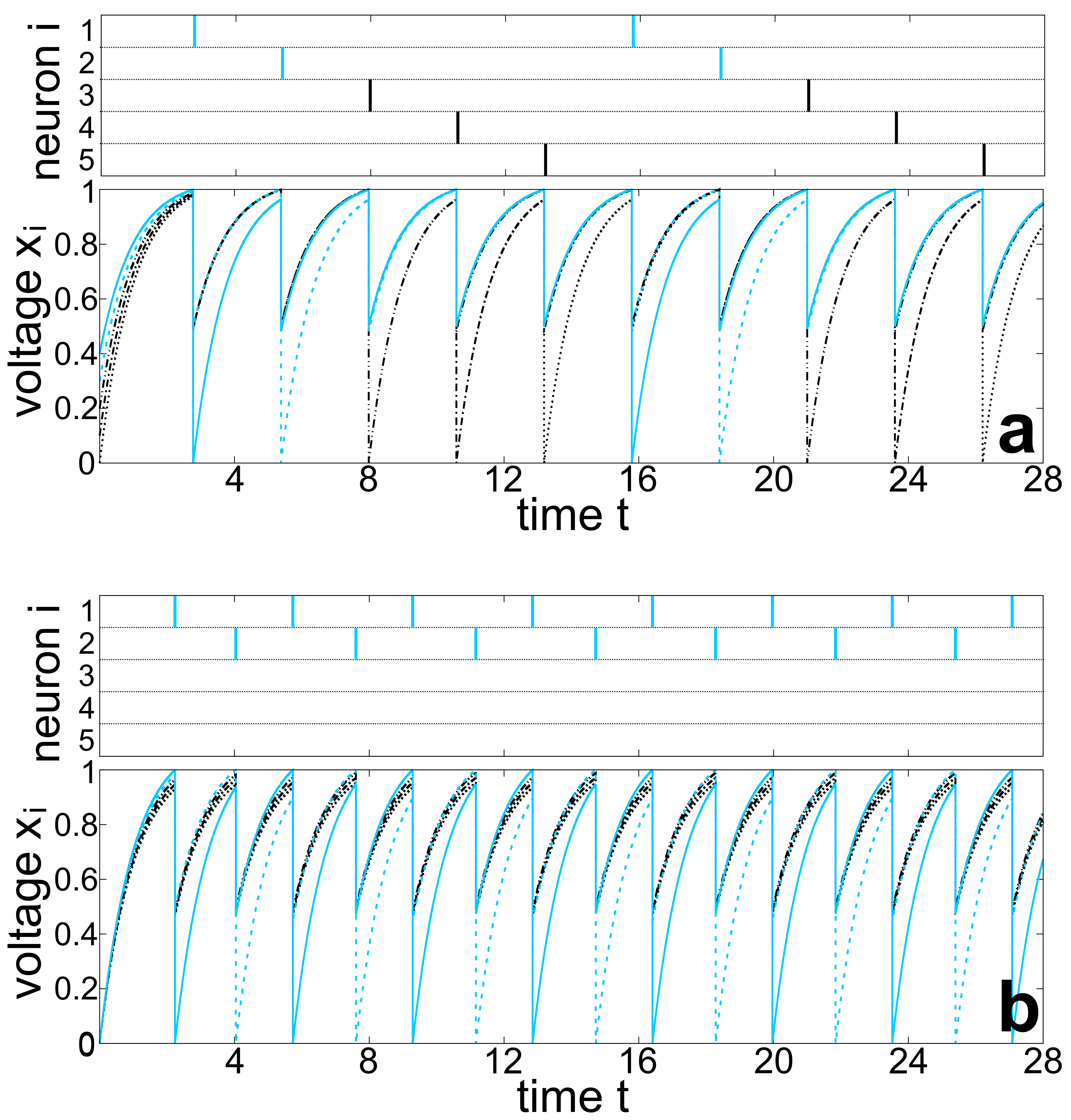}
\par\end{centering}
\caption[1st-entry]{ \small \textbf{Dynamics of 2-winners-take-all computation via pulse-suppression.} The voltage and spike times for all neurons in the network. Each curve represents the voltage dynamics of one neuron and each dash represents the time of an elicited spike. \textbf{(a)} shows the dynamics of all five neurons in the absence of external signals ($\xi(t)=(0,0,0,0,0)^{\intercal}$) such that all units spike periodically and sequentially. \textbf{(b)} Network dynamics when an asymmetric external signal $\xi(t) = (\xi_{1},\ldots,\xi_{5})^\mathsf{T}$ with $\xi_{1}>\xi_{2}>\xi_{3}>\xi_{4}>\xi_{5}$ is turned on. In the presence of external signals only neurons 1 and 2 spike (periodically and sequentially). Neurons 1 and 2 (blue curves) repeatedly overtake and inhibit neurons 3, 4 and 5 (black curves). Thus promoting 2-wta via pulse suppression. Parameters are $\epsilon = 0.5$ and $\Delta \xi = 0.02$.}
\label{fig:dynamics}
\end{figure}

How may spiking neural networks with pulse-suppression perform $k$-WTA functions? Its overall dynamics is dictated by a simple mechanism: every time a neuron reaches the firing threshold, it emits a spike and inhibits all other neurons proportionally to their voltages, thereby  bringing their state variables $x_i$ closer together, compare $d_2$ to $d_1$ in Fig.~\ref{fig:compression}. The neurons' voltages $x_i$ do not synchronize identically in finite time (unless they are initialized identically) because any resulting voltage  difference $x_i(t)-x_{i'}(t)$ is a positive fraction $(1-\epsilon)$ of the original $x_i(t^-)-x_{i'}(t^-)$. Furthermore, in the absence of external signals or noise, a neuron will also never overtake another \cite{Mirollo1990synchronization, Kielblock2011Overtaking}. This implies that all neurons repetitively reach the threshold sequentially one after the other and send a spike, see Fig.~\ref{fig:dynamics}a. The order in which they reach the threshold is determined by the ordering in their initial condition, i.e. their voltages at time zero. Thus, all neurons will exhibit  the same spiking frequency in the absence of external signals and noise.

% Figure 3 ends here

In the presence of noise (and still without external signals), sporadic overtaking becomes possible. Some neurons may by chance skip a spike emission (and a reset) due to inhibition. These events are random and the average number of spikes emitted by all neurons are still (almost) the same for any sufficiently large time interval, thus assuming identical average frequencies in the theoretical limit of infinite observation times.

\begin{figure}[]
\begin{centering}
\includegraphics[width=8.5cm,angle=0]{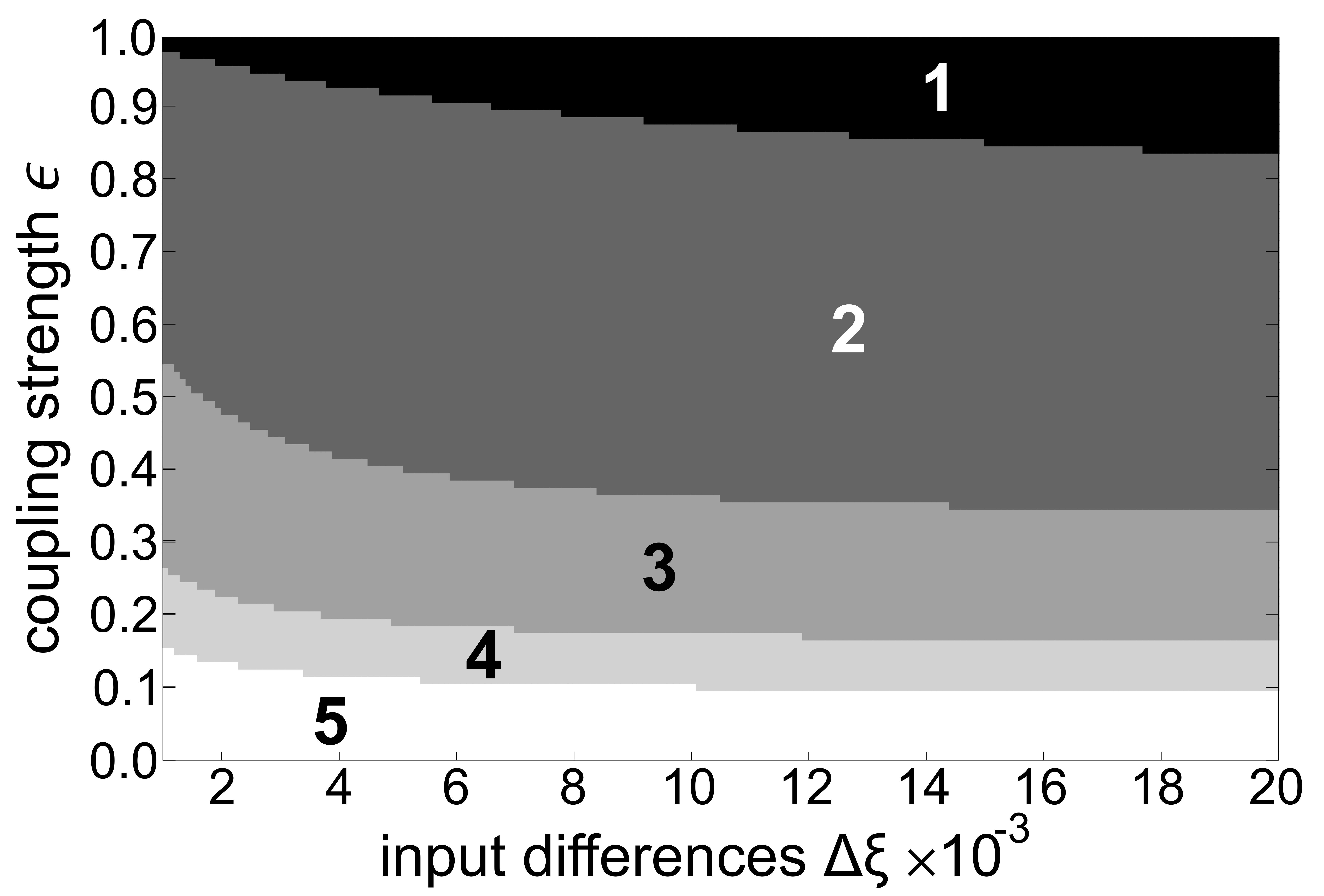}
\par\end{centering}
\caption[1st-entry]{ \small \textbf{(Re)configurable choice of computational task.} The computed $k$ depends on the differences $\Delta \xi$ and the coupling strength $\epsilon$. For a small network of five neurons, we observed five different response regions in the given parameter range: white, there is not ranking $k=5$; computing $k \in \{ 1,2,3,4\}$ are represented by different shades of gray or black. For any fixed input signal (horizontal value), a sorting type ($k$) can be selected by changing the global parameter $\epsilon$.}
\label{fig:function}
\end{figure}
% Figure 4 ends here

The external signals $\xi_i(t)$ that we define as inputs may induce much richer and more interesting collective network dynamics. If input signals with components of different intensities are concurrently applied to all neurons, neurons exhibit different intrinsic frequencies for the duration of the signal and the faster neurons may repeatedly overtake others, see Fig.~\ref{fig:dynamics}b. As a result, some neurons may never reach the threshold and deliver a spike, i.e. they will always skip their turn, (Fig.~\ref{fig:dynamics}b). We may thus interprete that the system evaluates some aspects of the external continuous-time analog input signal to make a decision about which spike emissions to skip and which to keep. As we will show below, the system performs a $k$-winners-take-all computation with $k$ depending on a combination of the network parameters and the magnitude of the difference between input signal's components.

\begin{figure*}[ht]
\begin{centering}
\includegraphics[width=15cm ,angle=0]{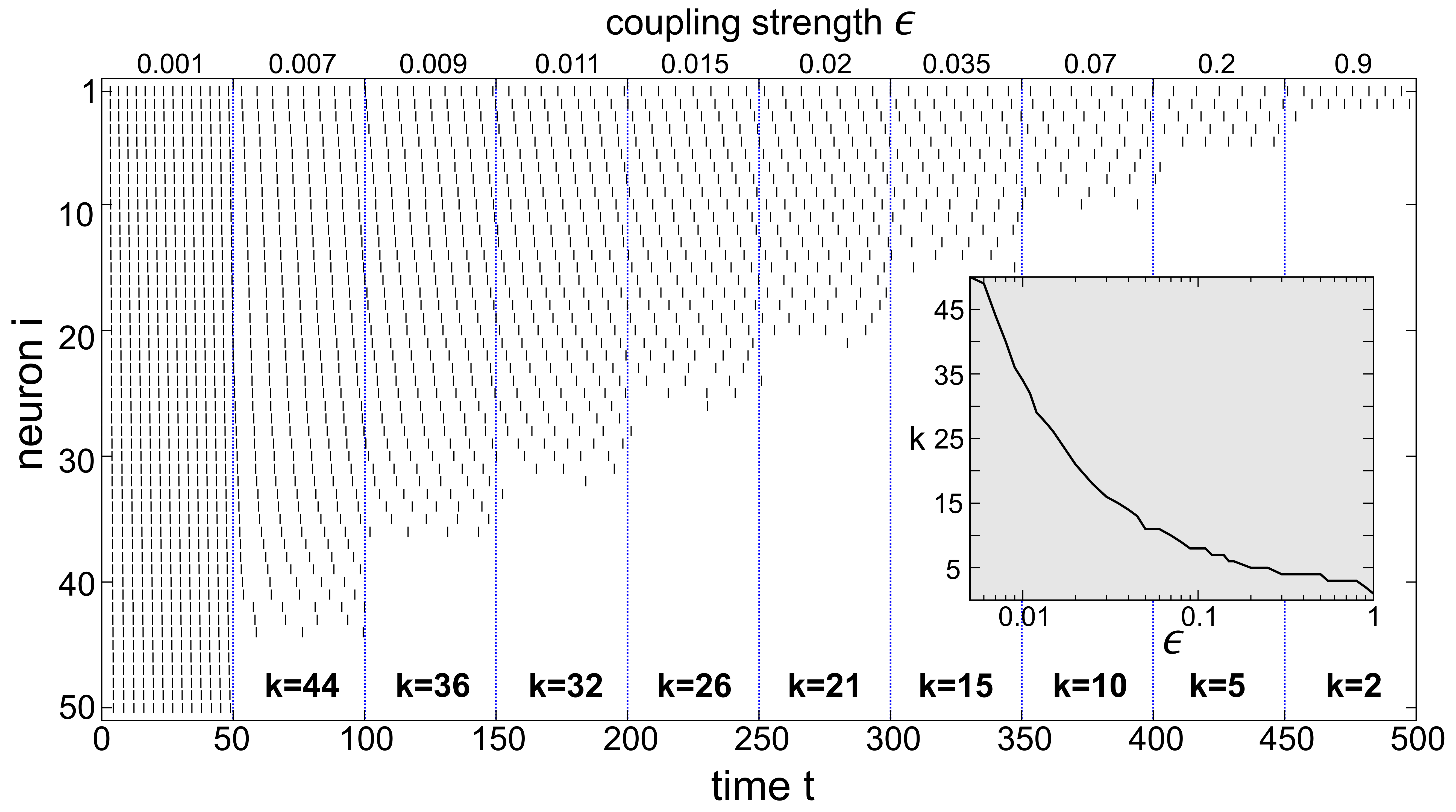}
\par\end{centering}
\caption[1st-entry]{ \small \textbf{ Monotonic decrease in computed $k$ with increasing $\epsilon$ for a network of $N=50$ neurons.} Each dash represents the time of one spike. For these simulations, inputs were fixed and linearly distributed between $0$ and $0.003$ ($\Delta\xi=6\times 10^{-5}$) and $\epsilon$ was switched to a new value every 50 time units. The inset shows how $k$ scales with $\epsilon$.}
\label{fig:50N}
\end{figure*}
% Figure 5 ends here

%\subsection{Reconfigurable computation}
Whether and how many neurons will skip their turn depends on two factors -- driving signals and coupling strength. First, we observe that if neurons are driven with input signals of different average values, neurons overtake each other (Fig.~\ref{fig:dynamics}b). If the input signals are constant in time, the  neurons $i$ receiving the larger driving signals $\xi_i$ exhibit faster voltage changes and are thereby candidates for being neurons with larger output spiking frequencies. We remark that even if a neuron exhibits a fast change in voltages it may not elicit any spike if inhibited, and the voltage is lowered by means of leakage rather than spikes delivered to the network. Details of their dynamics depend on how fast (or if at all) the neurons driven more strongly can catch up with more weakly driven ones ahead of them and overtake these. Second, the coupling strength $\epsilon$ determines how close all neurons' voltage variables are squeezed together during a spike reception. 

Whereas the first factor, different rates of change $dx_i/dt$ of the neurons' voltages, provides a simple mechanism for spreading the network voltages proportionally to their total input current into $N$ different voltage values, the effect of inhibitory spikes is more convoluted. It has two aspects: first, arriving spikes provide a synchronizing force for the group of $(N-1)$ neurons concurrently receiving the spike (Fig.~\ref{fig:compression}); second, whether the voltage difference from a neuron $j$ to the spike-sending neuron $i$ will increase or decrease depends on the phase $x_j(t_{s}^{-})$ at the spike time and $\epsilon$. For example, for $x_j (t_{s}^{-})$ close to one (almost synchronized) most $\epsilon$ increase the voltage difference $| x_i(t_s) - x_j(t_s) |$ as $x_i (t_s^-) \rightarrow 0$ and thus decrease the degree of synchrony; on the other hand, for initially large differences $| x_i(t_s^-)- x_j(t_s^-)|$, where $x_j(t_s^-)$ is near 0, all $\epsilon$ would increase synchrony. Furthermore, large values of $\epsilon$ provides a global synchronization force, e.g. if $\epsilon$ close to one all $x_j(t_s)$ are close to zero independent of $x_j(t_s)$. Moreover, whereas the first factor (different $dx_i/dt$) is acting continuously in time, the spike interaction is only acting at a discrete set of spike reception times. These observations imply that an intricate interplay of overall spiking frequency, differences in input signal strengths and coupling strength determines which type of output spike pattern results and how many neurons spike in the presence of a given input signal.

To quantify the contribution of these two factors, we simulate a system of $N=5$ neurons varying the global coupling strength $\epsilon$ and the degree of separation of input signal components. Specifically, we generate temporally constant signals of the form
\begin{equation}
\label{eq:input}
	\xi_{i} =  (N - i)\times \Delta \xi,  \qquad i\in \{1, \ldots, N\},
\end{equation}
where $\Delta \xi$ is a constant that characterizes the input signal. The smallest input difference between any two inputs therefore equals $\Delta \xi$. For all other system parameters fixed as above, we find that the system is capable of computing any given $k$, that is $k \in \{ 1,2,3,4,5 \}$ neurons spiking repeatedly whereas $N-k$ neurons remain silent without spikes. Fig.~\ref{fig:function} shows that for a single parameter $\epsilon$, the system will perform the same computation for a broad range of input differences $\Delta \xi$. Furthermore, for the entire range of $\Delta\xi$ considered the system can compute $k$-WTA for all $k \in\{1, \ldots ,5\}$, depending on the coupling strength $\epsilon$. The computation is thus reconfigurable in this sense. We remark that $\epsilon$ needs to be sufficiently large, otherwise no neuron is capable of overtaking another and thus $k=5$, i.e. no ranking is performed. Finally, in Fig.~\ref{fig:50N} we present a larger network example with $N=50$. As expected, we found that the mechanism described above applies in the same manner, and the global parameter $\epsilon$ controls the computed $k$. Specifically, $k$ decreases monotonically as $\epsilon$ increases. Furthermore, every $k \leq N$ is in principle possible for sufficiently well-tuned $\epsilon$.
\begin{figure*}[]
\begin{centering}
\includegraphics[width=15cm,angle=0]{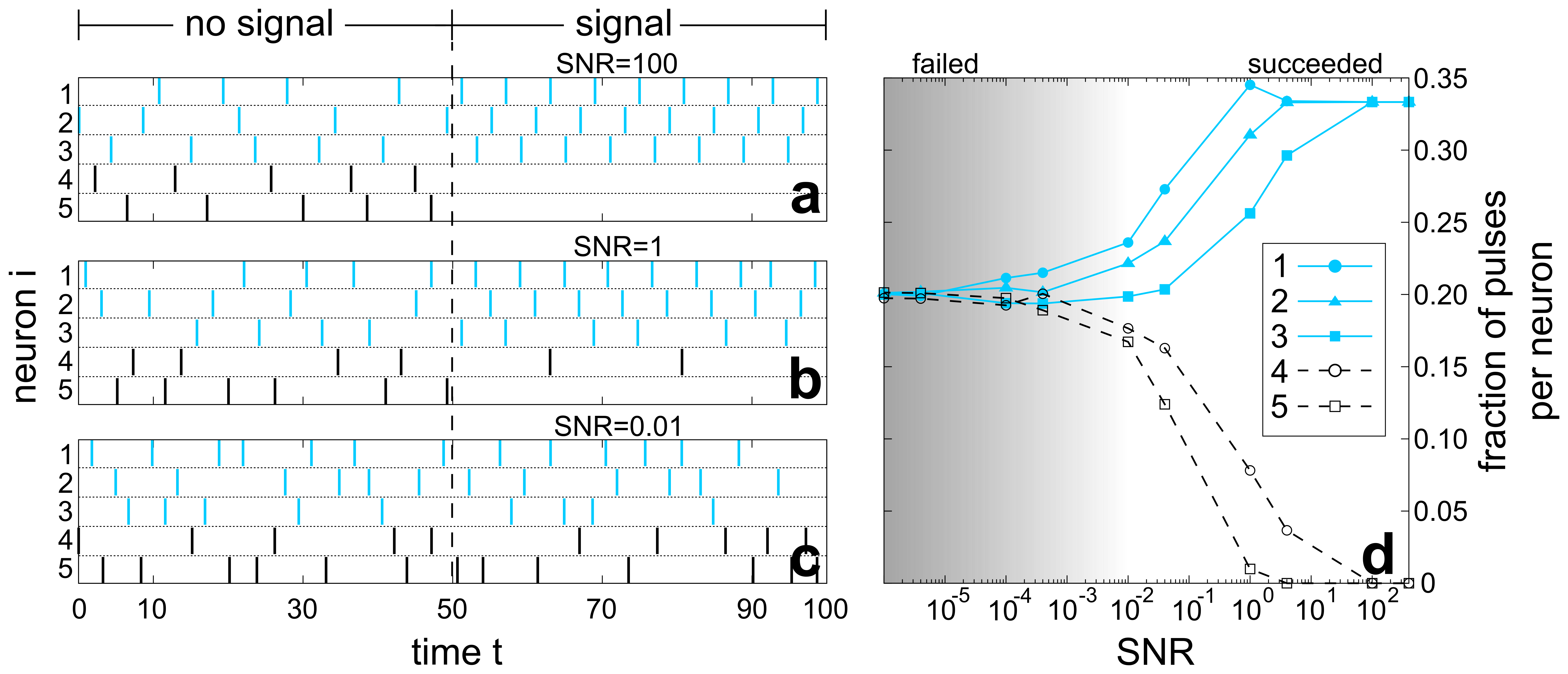}
\par\end{centering}
\caption[1st-entry]{ \small \textbf{Example of noise generating randomness in spike sequences and its stochastic effects.}. The system is set to compute the $k=3$ largest out of 5 inputs, as in (\ref{eq:input}).  \textbf{(a)} For very small noise ($\text{SNR} =100$), the system seems not affected by noise, as only the three stronger inputs induce spikes in this sample; \textbf{(b)} as noise increases ($\text{SNR} =1$), some predicted spikes are missed, replaced by false positives; \textbf{(c)} strong noise ($\text{SNR} =0.01$) generates a large number of false positives (spikes) and negatives, rendering the computation more difficult. \textbf{(d)} The fraction of spikes sent from each neuron in relation to the total network's number of spikes for different SNRs (single sample) counted for $t \in [0,2000]$. Even though the number of false positives increases with noise, the result of a computational task can still be solved via the average spike rate up to $\text{SNR} =0.01$. For example, for $\text{SNR}  \leq 0.001$ the system computes the wrong ranking and thus fails. Parameters used: $N=5$, $\epsilon=0.3$ and $\Delta\xi=0.003$}
\label{fig:noise}
\end{figure*}
% Figure 6 ends here

%\subsection{Noise and spike-rate variability}
As the voltages of many neurons may approach the threshold almost at the same time, see Fig.~\ref{fig:dynamics}, it is important to understand the concurrent effect of noise and input signals. We here consider a Gaussian white noise signal and inputs as in (\ref{eq:input}) for a $N=5$ network. In particular, we compute which fraction of all spikes comes from which neurons during a 3-WTA computation for different noise levels. To quantify noise in our system we define a signal-to-noise measure as
\begin{equation}
\label{eq:SNR}
\mbox{SNR}=\frac{\Delta\xi^2}{\sigma^2},
\end{equation}
where sigma is the standard deviation of the Gaussian signal and $\Delta\xi$ characterizes our signal strength as in (\ref{eq:input}). Let us consider the limiting cases first: in the absence of noise ($\sigma^2\rightarrow 0$), we expect that the two neurons subject to the weaker inputs would not elicit any spike while the other three neurons exhibit the same fraction ($\approx 33\%$) of spikes; for strong noise ($\sigma^2 \gg \Delta\xi$), we expect that most of the information about the input is lost, thus all neurons contribute roughly with the same $\approx 20\%$ of elicited spikes. As shown in Fig.~\ref{fig:noise}, our numerical results corroborate those predictions. Furthermore, as the signal to noise ratio (SNR) is reduced, the previously silent neurons start to send a fraction of the spikes inversely proportional to the SNR. This is particularly interesting, because it shows a form of robustness, that the computation can still be achieved up to SNRs approximately $0.01$ (Fig.~\ref{fig:noise}d), if the system computes with the average number of spikes, for example by imposing a threshold on the firing rates as a readout.

%\subsection{Computing Times with pulse-suppression}
So far, we have shown that the system introduced above is (re)configurable and how it responds to noise. Adjusting the coupling strength selects some $k$ for a $k$-WTA computation via changes in the spike rates. A fundamental feature of spiking systems it that a major part of consumed power is due to spike generation, i.e. the smaller the number of spikes the smaller the power consumed. Let us now briefly estimate how many spikes our system requires to compute at different SNR.

The number of spikes  $n^{*}$ required to rank the driving current from a pair of neurons, say $i$ and $j$ depends on how fast their individual numbers of generated spikes $n_i(t)$ and $n_j(t)$ diverge as function of time. Specifically, the difference $\Delta n_{ij} (t) = n_i(t) - n_j(t)$ encodes the rank result in its magnitude and sign, e.g. $\Delta n_{ij}(t)>0$ suggests that $\xi_i > \xi_j$ and $\Delta n_{ij}(t)<0$ indicates that $\xi_i < \xi_j$. We remark that the sign of $\Delta n_{ij}(t)$ alone is not enough to determine which neuron received the stronger input, due to variability caused by noise and initial conditions. We thus numerically estimate an expected failure rate function $F(\sigma,\Delta\xi, n(t))$ measuring how often $\Delta n_{ij} (t)$ indicates either the wrong signal order, e.g. $\Delta n_{ij} (t) > 0$ for $\xi_i < \xi_j$, or is inconclusive ($\Delta n_{ij} (t) = 0$) after a total of $n(t)$ network spikes, calculated over 500 simulations per parameter set. The minimum number $n^*$ of spikes from the whole network to compute the partial rank order with an acceptable failure rate $\delta$, noise standard deviation $\sigma$ and input amplitude $\Delta \xi$, is then estimated as
\begin{equation}
n^{*}(\sigma, \Delta \xi, \delta) = \min \{n^\prime \, | \, F(\sigma, \Delta \xi, n) < \delta \, \text{ for all } \, n > n^\prime\}.
\label{eq:tempo}
\end{equation}
The quantity $n^*$ thus define the minimum (total) number of spikes after which the failure rate will never exceed $\delta$ afterwards within a much larger number of spikes, interval $[0,2000]$.
\begin{figure*}[]
\begin{centering}
\includegraphics[width=15cm,angle=0]{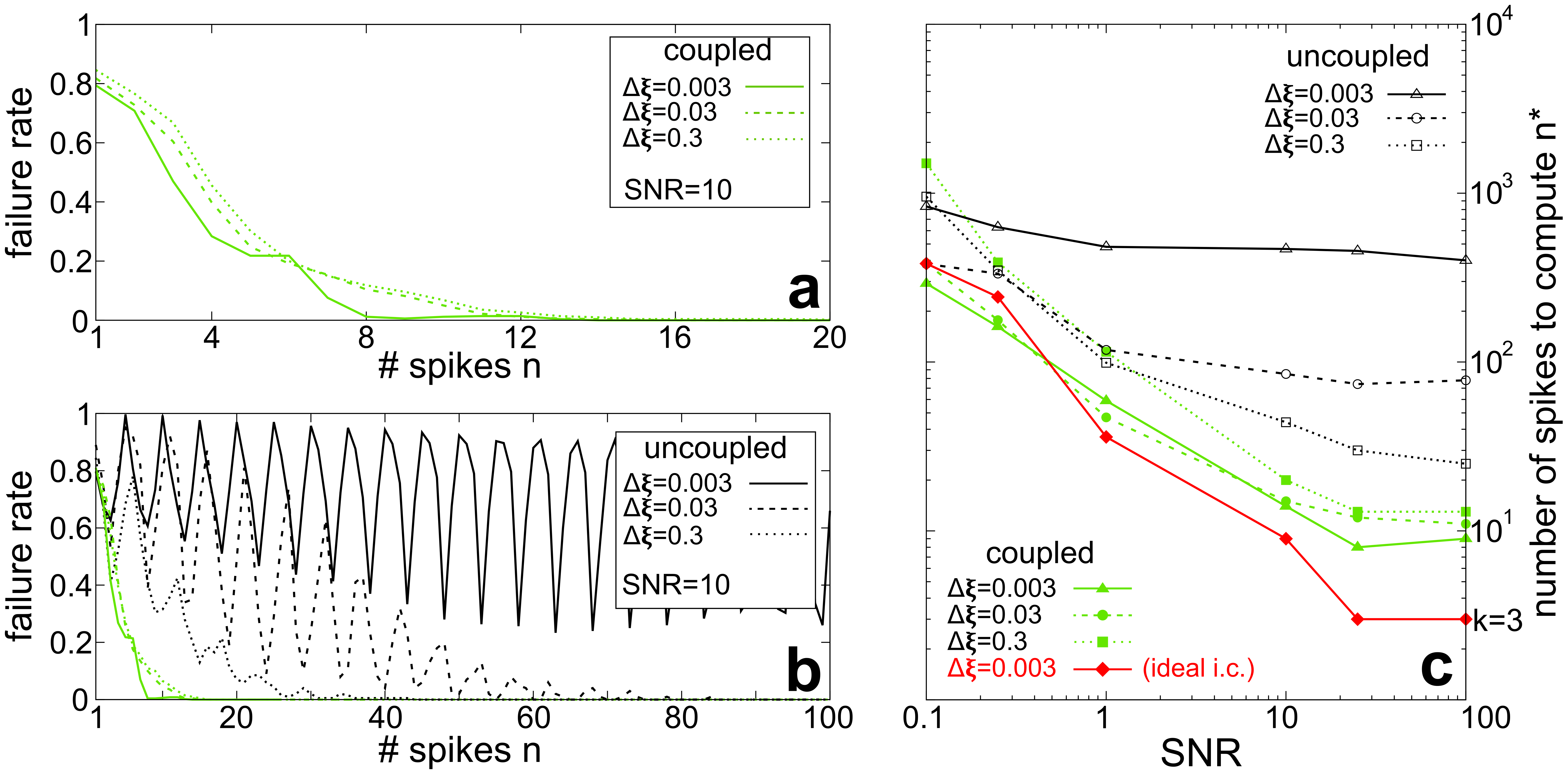}
\par\end{centering}
\caption[1st-entry]{ \small \textbf{Number of spikes required to compute with a coupled and an uncoupled systems.} \textbf{(a)} For small SNR, the coupled system computes with few spikes, consistent across different input magnitudes and fixed SNR, thus is robuts against $\Delta\xi$ changes. \textbf{(b)} the uncoupled system requires a large variety of number of spikes ($n^*$) for the same SNR, depending on the input strength (results for coupled system in green). Because all $N=5$ neurons have similar spike frequencies, an oscillation with period close to 5 is observed, as $\Delta n_{ij}$ has a larger probability to be zero for all $i,j$ pairs at times in which $n(t)$ is a multiple of five (one spike per unit). Deviations on this period are cause by spike frequency variability (overtakes). \textbf{(c)} The coupled system consistently exhibits a smaller number of spikes and a smaller variation at the same SNR when compared to the uncoupled system. At small SNRs and ideal initial conditions ($V_i=0$ for all $i$), in red, only the absolute minimum number of spikes 3 (n=k) is required. Parameters used are $\epsilon=0.3$ and $\delta=0.002$.}
\label{fig:speed}
\end{figure*}
% Figure 7 ends here

As an example, we consider a $N=5$ network set to compute a 3-WTA ranking over five inputs chosen as in (\ref{eq:input}). In this experiment $k=3$ is known and we are interested on comparing the three winners to the two losers. Among those comparisons the smallest difference among input strengths is $\xi_3 -\xi_4 = \Delta \xi$, between neurons 3 and 4. Their difference in spike frequency determines the overall computing time because it dictates the longest relevant computing time in the process. Notice that the in-group differences are irrelevant for the desired computation. Moreover, the in-group rankings may take longer than determining the three stronger inputs and not possible to compute for large SNRs. For reference, we compare our results to a set of independent integrators. For clarity, we consider the exact same model of neurons described above, but now uncoupled, thus stressing the computational aspects provided by the network dynamics rather than model specific features.

Let us first consider the limit case of no noise ($\sigma \rightarrow 0$) and an ideal initial condition where all neurons start at the same voltage $V_i(0) = 0$. In this limit, in the coupled system, neuron 4 does not spike and the number of spikes to compute is simply $n=3$, one spike from each winner, and decoding the resulting spike trains reduces to detecting the first three spikes. On the other hand, the number of spikes to compute with the uncoupled system is  proportional to $\Delta \xi$ and is given by the $n(t)$ in which $|\Delta n_{34}(t)|>2$, with no actual limit to how large the number of spikes may become (minimum of 2). In the other extreme limit of very large noise, $\Delta n_{34}(t)$ is approximately a discrete random walk with step size one and zero mean, thus no computation is performed in either system. 

More interestingly, Fig.~\ref{fig:speed} illustrate the intermediary cases where noise may (or may not) promote spike variability on all five neurons. In particular it shows how the number of spikes needed to compute scales with the SNR in both coupled and uncoupled systems. We found that not only the coupled system consistently computes with less spikes across a broad range of SNRs but also exhibits a small variability when the input magnitude is varied (Fig.~\ref{fig:speed}a). Furthermore, for ideal initial conditions, $V_i(0)=0$ for all $i$, the coupled system computes with the absolute minimum number of spikes $n^*=3$ for $\text{SNR} \geq 25$. In contrast, the uncoupled system (Fig.~\ref{fig:speed}b) does not exhibit a predictable number of spikes $n^*$ to compute as a function of the SNR, as it is also a function of the input signal differences $\Delta\xi$, and presents a spike response variability with orders of magnitude of difference for the same SNR (Fig.~\ref{fig:speed}c). In practice, given an interval of possible SNRs and input strengths, the largest estimated $n^*$ should be chosen as an effective number of spikes to finish a computation.

The increase in computing speed due to reduced noise observed on the coupled system occurs because it is itself computing the rank order as it goes, i.e. the neurons are not only mapping the input strength to spike trains with different frequencies but actively inhibiting other components in a self-organized way. In summary, 	$k$-WTA by pulse-suppression is computed with reasonably few spikes ($3 \leq n^* \leq 70$, numerically estimated) if the SNR is kept relatively large ( $\geq 1$) and for $\text{SNR}  \geq 25$ only the absolute minimum number of spikes $n^*=k$ is required, thus the computing time is only limited by the neurons' intrinsic spike frequency. As the mechanism of self-organized pulse suppression is independent of both model details and network size, these results should hold qualitatively to other models and network sizes.

\section{Discussion} % SECTION 4: DISCUSSION
Our results show that fast-computing (re)configurable $k$-WTA can be implemented by simple symmetrical systems of spiking neurons with only inhibitory interactions, providing a versatile model capable of performing a variety of functions using the same underlying network structure. We believe our model bridges and complements a variety of other approaches to $k$-WTA that trade versatility, e.g. reconfigurability and/or simple topology, for specific features as output stability and sensitivity to weak input signals. For example Heteroclinic Computing \cite{Neves2009controlled,Neves2012computation,Neves2017Noise} offers sensitivity to arbitrarily small signals (up to the noise level), because it relies only on unstable states, while neural fields \cite{Sandamirskaya2014dynamics, Strub2017dynamic} provide a soft-WTA with macroscopic stability (population dynamics) via short-range excitation and long-range inhibition. Our model makes use of stable orbits and can robustly compute in a given time over a broad range of input magnitudes. Moreover, we expect the results on the influence of noise to stay qualitatively the same across other types of noise as long as they exhibit zero mean and exhibit no long-term (power law) correlations. Such types would add a bias proportional to the mean amplitude because noise is simply integrated in-between pulse-events.

Due to its simplicity, our model may provide the basis for novel hardware implementations. That is, our neuronal model is one-dimensional and generic, thus has a variety of well-established implementations, while the amplitude of interacting spikes is simply proportional to the voltage of the neuron receiving them, resembling a simple leakage of current to a common ground. Furthermore, the network topology employed is symmetrical, so it does not rely on any specifically weighted or complex topological interactions. We remark that system symmetry is not a necessary condition for such types of computations even though our theoretical model represents ideal symmetrical components. Small variations in parameters, e.g. coupling weights and neuron spiking frequencies, would add a bias to the results with strength proportional to the given variations because after integration all elements on the right-hand side of Equation~(\ref{eqn:voltageODE}) represent simple currents.

As a versatile model capable of, in principle, performing any $k$-WTA computation for a given network size $N$ and using only few spikes, it may be a good candidate for implementing functionality on autonomous systems. Varying $k$ means that a small network is already capable of performing many different functions while the low spike count may translate in low power approaches to computation. Thus, our results may potentially also contribute to the growing field of artificial cognitive computing and related topics \cite{memmesheimer2006designing,memmesheimer2006designing2,schoner_2008, lukovsevivcius2009reservoir, Steingrube2010pattern, Hopfield2015understanding,abbott2016FunctionalNetworksSpikingNeurons, Strub2017dynamic}.

\section*{Acknowledgement}
This work was supported in part by the German Science Foundation (DFG) under Germany's Excellence Strategy -- EXC-2068 -- 390729961- Cluster of Excellence Physics of Life of TU Dresden, the DFG and the Free State of Saxony by a grant towards the Research Cluster Center for Advancing Electronics Dresden (cfaed), the Free State of Saxony towards the project TransparNET and by the DFG under grant number TI 629/5-1.

%%%%%%%%%%%% BIBLIOGRAPHY %%%%%%%%%%%%%%%
\bibliographystyle{unsrt}
\bibliography{ReconfigurableComputation2020}

\end{document}